\documentclass[3p,times,procedia]{elsarticle}
\usepackage{nupha_ecrc}

\volume{00}

\firstpage{1}

\journalname{Nuclear Physics A}

\runauth{C. Shen}

\jid{nupha}

\jnltitlelogo{Nuclear Physics A}

\usepackage{amssymb}

\biboptions{sort&compress}

\usepackage[figuresright]{rotating}

\begin{document}

\begin{frontmatter}



\dochead{XXVIIIth International Conference on Ultrarelativistic Nucleus-Nucleus Collisions\\ (Quark Matter 2019)}

\title{Studying QGP with flow: A theory overview}


\author[add1,add2]{Chun Shen}

\address[add1]{Department of Physics and Astronomy, Wayne State University, Detroit, MI 48201, USA}
\address[add2]{RIKEN BNL Research Center, Brookhaven National Laboratory, Upton, NY 11973, USA}

\begin{abstract}
I review recent developments in the phenomenological study of the quark-gluon plasma (QGP) transport properties based on a personal selection of results that were presented at Quark Matter 2019.
The constraints on the temperature dependence of QGP shear and bulk viscosity are summarized. I discuss new theory advancements towards more realistic 3D dynamical simulations of heavy-ion collisions at finite baryon density. The challenges and opportunities of applying hydrodynamics to small collision systems are highlighted.
\end{abstract}


\end{frontmatter}


\section{Introduction}
\label{Intro}

Relativistic heavy-ion collisions create extreme conditions, under which quarks and gluons are in a deconfined phase, the quark-gluon plasma (QGP). The QGP exhibits emerging collective behavior that stems from the many-body interactions in Quantum Chromodynamics (QCD). The nearly perfect fluidity of QGP has been investigated extensively in heavy-ion experiments, in which an unprecedented level of precision in the experimental observables has been achieved. The harmonic flow coefficients $V_n = v_n e^{i n \Psi_n} \equiv \sum_j e^{i n \phi_j}$ are introduced to represent the collective anisotropic particle distribution in heavy-ion collisions \cite{Ollitrault:1992bk}. Event-by-event fluctuations and the pressure-driven QGP evolution are encoded in the correlation measurements of the flow magnitude $v_n$ and their phases $\Psi_n$, such as the cumulants of harmonic flow $v_n\{m\}\,(m = 2, 4, 6, 8, ...)$ \cite{Khachatryan:2015waa}, symmetric cumulants \cite{ALICE:2016kpq}, event-plane correlators \cite{Aad:2014fla}, and event-by-event flow distributions \cite{Aad:2013xma}. 
In the last decade, relativistic hydrodynamics, including lattice QCD based equation of state (EoS), viscosity, and initial state fluctuations, has been developed to understand the dynamics of QGP evolution and the flow observables. Such a theoretical framework has successfully described and even predicted various types of flow correlation measurements with remarkable precision.
Up-to-date, phenomenological studies of flow observables have been a precision tool to elucidate the QCD phase structure, the equation of state of hot and dense nuclear matter, the QGP viscosity, conserved charge diffusion, and event-by-event quantum fluctuations in heavy-ion collisions (see reviews \cite{Heinz:2013th,Gale:2013da}). Finally, the dynamical evolution of the QGP plays an indispensable role in understanding electromagnetic radiation (direct photons and dileptons) \cite{Shen:2016odt} and strongly-interacting probes, such as QCD jets and heavy quarks \cite{Putschke:2019yrg}. 

\section{State-of-the-art extraction of the QGP transport properties: A 2019 status report}
\label{viscosity}
\vspace{-2mm}

Fluid dynamics emerges as an effective description of the system’s evolution when the mean free path is small compared to macroscopic scales like the system size. The equations of motion of hydrodynamics are the conservation laws of energy, momentum, and conserved charge currents (namely net baryon, strangeness, and electric charges) supplemented by relaxation equations for dissipative fluxes.
The physical properties of the fluid can be characterized by thermodynamic relations from the equation of state and several transport coefficients, such as the specific shear and bulk viscosity. Israel and Stewart’s formulation of relativistic dissipative fluid dynamics \cite{Israel:1979wp} ensures causality and stability and has been the main theory applied to describe the time evolution of the QGP at RHIC and the LHC.
See Refs. \cite{Baier:2007ix,Denicol:2012cn,Strickland:2014pga,Jeon:2015dfa} for details about derivations of transient fluid dynamics from various types of microscopic theory. 
The fluid-dynamic description naturally breaks down as the system becomes increasingly dilute within its hadronic phase.
One must then transit to a microscopic transport description. The numerical realizations of hadronic transport models are UrQMD \cite{Bass:1998ca,Bleicher:1999xi}, JAM \cite{Nara:1999dz}, and SMASH \cite{Weil:2016zrk}.

\begin{figure}[t!]
    \centering
    \includegraphics[width=0.98\textwidth]{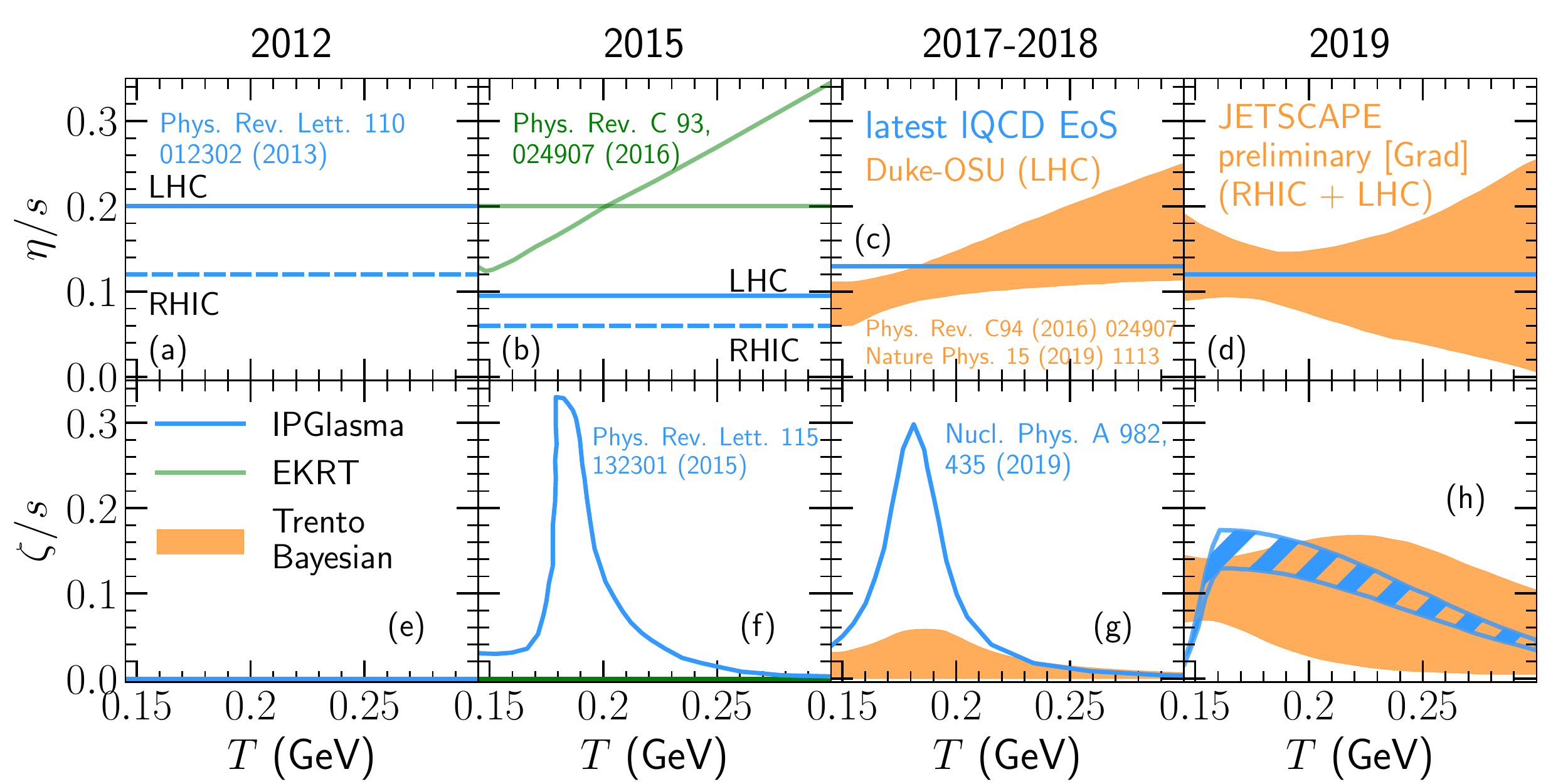}
    \caption{A summary of phenomenological constraints on the temperature-dependent QGP specific shear and bulk viscosity from 2012 to 2019. The progressing constraints on the QGP viscosity used in the frameworks with the IP-Glasma initial conditions \cite{Gale:2012rq, Ryu:2015vwa, Schenke:2018fci, Schenke:2020unx, Gale:QM2019} are compared with those using other types of initial state models. The result using the EKRT initial condition is from Ref. \cite{Niemi:2015qia}. The Bayesian extracted QGP $(\eta/s)(T)$ and $(\zeta/s)(T)$ \cite{Bernhard:2016tnd,Bernhard:2019bmu} was constrained by flow measurements in p+Pb and Pb+Pb collisions at the LHC. The orange bands indicate a 90\% confidence level. In this conference, the preliminary Bayesian analysis calibrated with combined RHIC and LHC flow measurements was presented by the JETSCAPE Collaboration \cite{Paquet:QM2019}. The blue hatched area in (h) indicates the variation of $(\zeta/s)(T)$ with (upper) and without (lower) the EKT phase.}
    \label{fig1}
\end{figure}

In this conference, a start-of-the-art theoretical framework was presented to perform multi-stage dynamical simulations of heavy-ion collisions \cite{Gale:QM2019}. It consists of the event-by-event IP-Glasma initial state \cite{Schenke:2012wb} connected with a pre-hydrodynamic phase modeled by an effective QCD kinetic theory (EKT), K{\o}MP{\o}ST \cite{Kurkela:2018wud}. After the latter drives the collision system sufficiently near local equilibrium, the full energy-momentum tensor is fed into second-order viscous hydrodynamics (MUSIC) \cite{Paquet:2015lta} and finally transition to a hadronic transport, UrQMD \cite{Bass:1998ca,Bleicher:1999xi}, in the dilute density phase. Remarkably, this framework can quantitatively reproduce a variety of flow measurements in heavy-ion collisions from 200 GeV to 5020 GeV \cite{Gale:QM2019} with an effective $(\eta/s)_\mathrm{eff} = 0.12$ and a temperature-dependent bulk viscosity (see Fig.~\ref{fig1}h). A direct comparison with the simulations without the EKT phase \cite{Schenke:2020unx} showed that the conformal EKT generates more flow than viscous hydrodynamics at this early stage. A similar finding was earlier observed for free-streaming dynamics \cite{Liu:2015nwa}. The additional EKT phase thus leads to a $35\%$ larger extracted QGP bulk viscosity. That study demonstrated the significant phenomenological impact of a realistic modeling of the early stage of heavy-ion collisions on constraining the QGP bulk viscosity. 

To systematically constrain the QGP transport properties, adopting the Bayesian statistical analysis has become a standard approach in our field \cite{Pratt:2015zsa, Bernhard:2019bmu}. In this conference,  the JETSCAPE Collaboration presented a systematic Bayesian analysis, which was calibrated with combined flow measurements from RHIC and the LHC for the first time \cite{Paquet:QM2019}. This work demonstrated that a simultaneous calibration using flow observables at two collision energies, which differ by an order of magnitude, led to strong constraints on the temperature dependence of the QGP shear and bulk viscosities. Heavy-ion collisions at RHIC and the LHC offer a wide dynamical range for the theoretical framework to explore the parameter space. At the same time, different types of off-equilibrium corrections at the particlization stage introduce sizable theoretical uncertainties to the Bayesian extraction, demanding more theoretical work in this direction. The JETSCAPE Collaboration also emphasized that performing closure tests in Bayesian analysis was an essential step before extracting any physical information from experimental measurements \cite{Paquet:QM2019}.

Fig.~\ref{fig1} summarizes the collective effort of quantifying the QGP transport properties over the past seven years in our field. As theoretical tools are being developed rapidly to include more and more physics, the extraction of QGP transport properties becomes more systematic. 
The saturation-based IP-Glasma and EKRT initial conditions prefer an effective shear viscosity between 0.12 and 0.20 in the hydrodynamic phase to achieve a simultaneous description of all orders of harmonic flow coefficients \cite{Gale:2012rq,Niemi:2015qia}.
Owing to large pressure gradients and finite initial radial flow in the IP-Glasma initial conditions, a temperature-dependent bulk viscosity is essential to balance the consequent strong flow and reproduce the mean $p_T$ measurements in the heavy-ion collisions. At the same time, the introduction of the bulk viscosity reduced the extracted QGP shear viscosity by almost 50\% \cite{Ryu:2015vwa}. By adopting the equation of state from the latest lattice QCD calculations \cite{Borsanyi:2013bia,Bazavov:2014pvz}, the Duke-OSU group delivered the first Bayesian Inference on the temperature-dependent shear and bulk viscosities \cite{Bernhard:2016tnd,Bernhard:2019bmu}. There was tension of the Bayesian extracted bulk viscosity $(\zeta/s)(T)$ \cite{Bernhard:2016tnd,Bernhard:2019bmu} with the parameterization used in the IP-Glasma hybrid framework \cite{Schenke:2018fci}. This difference was greatly reduced in 2019. On the one hand, the significant changes in the IP-Glasma hybrid framework come from allowing the peak temperature of bulk viscosity to drop from $T_\mathrm{peak} = 180$ MeV to $160$ MeV. A lower $T_\mathrm{peak}$ with a smaller $(\zeta/s)_\mathrm{max}$ is favored by hadron mean $p_T$ measurements in peripheral Pb+Pb and Au+Au collisions, in which the maximum temperatures at the starting time of hydrodynamic simulations are close to, or even below, $180$ MeV. On the other hand, a more flexible prior parameterization of $(\zeta/s)(T)$ in the JETSCAPE Bayesian analysis allows a large QGP bulk viscosity in the posterior distribution for the model to reproduce flow measurements.
Lastly, it is worth emphasizing that all the theoretical frameworks used in Fig.~\ref{fig1} are open-source and publicly available to the entire community. 

Other statistical methods, such as Principal Component Analysis (PCA), are applied to study the fluctuations and correlations in heavy-ion collisions \cite{Mazeliauskas:2015efa,Hippert:2019swu,Gardim:2019iah}. See the review by L. Pang for more details about applying machine learning techniques to heavy-ion collisions \cite{Pang:QM2019}.

\vspace{-5mm}
\section{Theoretical modeling of heavy-ion collisions at $\mathcal{O}(10)$ GeV: The era of full 3D dynamics}
\label{BES}
\vspace{-2mm}

\begin{figure}[ht!]
    \centering
    \includegraphics[width=0.98\textwidth]{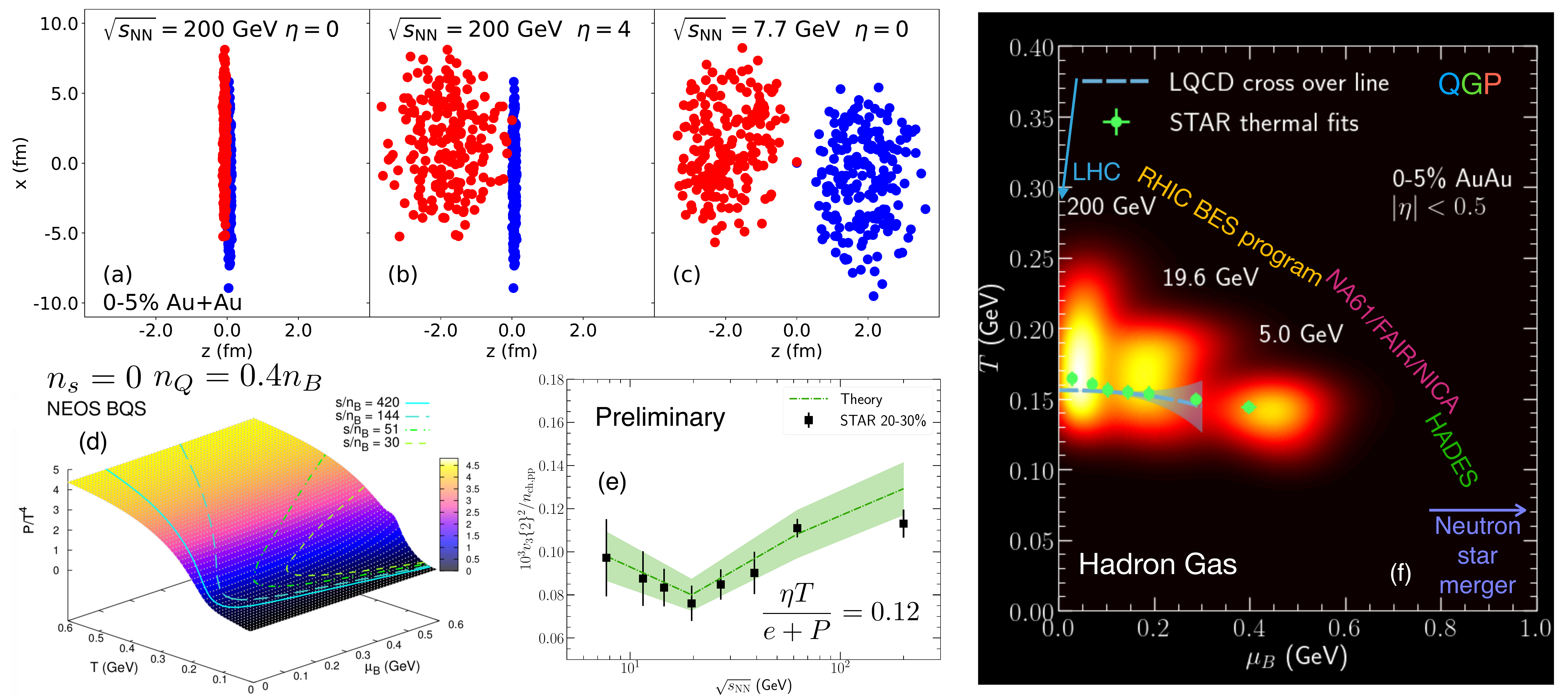}
    \caption{Relativistic heavy-ion collisions at the RHIC Beam Energy Scan program explore the nuclear matter phase diagram and study QGP at finite net baryon density. The panels (a-c) show longitudinal Lorentz contraction of the two colliding nuclei at two rapidities and collision energies \cite{Shen:2017bsr}. \textit{Panels (d):} The equation of state of QCD matter at finite densities \cite{Monnai:2019hkn}. \textit{Panels (e):} Preliminary theoretical calculation of the collision energy dependent triangular flow compared with the STAR measurements \cite{Adamczyk:2016exq}.
    \textit{Panel (f):} Our current knowledge about the QCD phase diagram with current and future experimental heavy-ion programs. Lattice QCD calculations identified a smooth cross-over between QGP and hadron gas for $\mu_B < 250$ MeV \cite{Bazavov:2018mes}. The chemical freeze-out points extracted from thermal fits at RHIC \cite{Andronic:2017pug,Adamczyk:2017iwn} are shown. The three blobs represent fireball trajectories of Au+Au collisions at RHIC BES energies mapped onto the QCD phase diagram event-by-event using a dynamical framework \cite{Shen:2017bsr}. Their brightness is proportional to the fireball space-time volume weighted by $T^4$.}
    \label{fig2}
\end{figure}

Current heavy-ion experiments in the RHIC Beam Energy Scan (BES) program, the NA61/SHINE experiment at the Super Proton Synchrotron (SPS), and future experiments at the Facility for Antiproton and Ion Research (FAIR) and Nuclotron-based Ion Collider fAcility (NICA) produce nuclear matter to probe an extensive temperature and baryon chemical potential region in the phase diagram.
Precise flow measurements of the hadronic final state ought to allow for the extraction of transport properties of the Quark-Gluon Plasma (QGP) in a baryon-rich environment. In order to do so, we require a reliable theoretical framework that can model the dynamical evolution of the collisions and all relevant sources of fluctuations. Fig.~\ref{fig2}f presents our current knowledge of the QCD phase diagram. The recent calculation of higher-order susceptibilities from lattice QCD provides us with a crossover line up to $\mu_B \sim 250-300$ MeV \cite{Bazavov:2018mes}. The estimated crossover line agrees with the chemical freeze-out temperatures and net baryon chemical potentials extracted from the STAR BES hadron yield measurements \cite{Andronic:2017pug,Adamczyk:2017iwn}. The three blobs indicate the fireball trajectories for typical AuAu collisions at 200, 19.6, and 5 GeV. There has been an increasing interest in understanding how the QGP transport properties vary as the system is doped with more and more baryon charge. The QGP in such a baryon rich environment also allows us to study conserved charge diffusion.

Heavy-ion collisions at $\sqrt{s} \sim \mathcal{O}(10)$ GeV strongly violate longitudinal boost invariance and require full 3D modeling of their dynamics. As the collision energy decreases or in forward rapidity regions, the reduced Lorentz contraction along the beam direction leads to a significant overlap time for the two nuclei to pass through each other (see illustrations in Figs.~\ref{fig2}a-c). Therefore, understanding pre-equilibrium dynamics has become an active topic. Initial state models based on classical string deceleration \cite{Shen:2017bsr,Bialas:2016epd} and hadronic transport approaches \cite{Karpenko:2015xea,Du:2018mpf} have been integrated into the full theoretical framework to model the dynamics of heavy-ion collisions at low energies. There are recent new theoretical developments to understand early-stage baryon stopping from the CGC-based approaches in the fragmentation region \cite{Li:2018ini,McLerran:2018avb} and a holographic approach at intermediate couplings \cite{Attems:2018gou}.
Having an extended interaction region in heavy-ion collisions requires treating the energy-momentum and charge density currents as sources in the hydrodynamic equations of motion. By feeding them gradually into the hydrodynamic fields on a local basis while the two nuclei pass through each other we can interweave the initial collision dynamics with hydrodynamics. Such a dynamic initialization scheme was proposed \cite{Okai:2017ofp,Shen:2017ruz} and has been adopted by several groups \cite{Shen:2017bsr,Du:2018mpf}. In this conference, Y. Kanakubo presented a dynamical initialization framework to achieve a uniform description of the hadronic chemistry from pp to AA collisions \cite{Kanakubo:2019ogh}. 

Hydrodynamic simulations at low energies require an equation of state (EoS) for the nuclear matter at finite net baryon density, which cannot be directly computed from lattice QCD because of the sign problem \cite{Ratti:2018ksb}. Based on the current knowledge of higher-order susceptibilities at $\mu_B = 0$ MeV, estimated EoS can be constructed through a Taylor expansion \cite{Monnai:2019hkn, Noronha-Hostler:2019ayj, Parotto:2018pwx}. Fig.~\ref{fig2}d highlights an equation of state at finite net baryon density assuming strangeness neutrality and electric charge density $n_Q = 0.4 n_B$. The enforcement of strangeness neutrality improves the description of relative particle yields for multi-strange particles measured in Pb+Pb collisions at the top SPS energy \cite{Monnai:2019hkn}.
The progress described above enables event-by-event dynamical simulations for heavy-ion collisions at the RHIC BES program. At this conference, a fully integrated framework \cite{Shen:2017bsr,Shen:2018pty,Monnai:2019hkn} was shown to reproduce the rapidity dependence of particle production as well as the collision energy dependence of the STAR $v_n\{2\}\,(n = 2,3)$ flow measurements in Au+Au collisions from 200 GeV to 7.7 GeV \cite{Adamczyk:2017hdl}.  Remarkably, this preliminary calculation highlighted in Fig.~\ref{fig2}e can produce a similar non-monotonic collision energy dependence present in the experimental triangular flow data measured at the RHIC BES phase I \cite{Adamczyk:2016exq}, without the need for a critical point in the phase diagram. Therefore, it is essential to understand the interplay among the duration of dynamical initialization, the variation of the speed of sound, and the $T$- and $\mu_B$-dependence of the specific shear viscosity.
That work demonstrated a critical role of theoretical modeling in elucidating the origin of the non-monotonic behavior seen in the RHIC measurements. Phenomenological studies of the precise anisotropic flow measurements from the upcoming analysis of RHIC BES II will be able to further constrain the $\mu_B$ dependence of the QGP shear and bulk viscosities. 

Flow correlations between different rapidity regions reflect event-by-event longitudinal fluctuations in the full 3D dynamics \cite{Bozek:2017qir, Li:2019eni, McDonald:2020oyf}. The anisotropic flow coefficients decorrelate as a function of particle rapidity. A systematic study by A. Sakai showed that initial state fluctuations and thermal fluctuations in hydrodynamics were equally important to understand the centrality dependence of flow decorrelation measurements at LHC \cite{Sakai:QM2019}. One theory calculation based on the 3D-Glauber initial conditions \cite{Shen:2017bsr} over-predicted the collision energy dependence of the flow decorrelation from 200 down to 27 GeV measured at the RHIC BES program \cite{Nie:QM2019}. Measurements from a wide range of collision energy thus have strong constraining power for our understanding of the longitudinal dynamics in heavy-ion collisions.

Finally, the dynamics of conserved charge density currents inside the QGP can elucidate the medium's charge diffusion constants and heat conductivity, which are so far poorly constrained transport properties of the QGP.
This topic has been stimulating interest in developing realistic initial conditions for conserved charge distributions \cite{Martinez:2019jbu,Martinez:2019rlp} and detailed modeling of the QGP chemistry \cite{Pratt:2012dz}.
In this conference, the longitudinal dynamics of coupled multiple charge diffusion processes were presented by J. Fotakis for the first time \cite{Greif:2017byw, Fotakis:2019nbq}.
The baryon diffusion transports baryon charges from forward rapidity back to the central rapidity region. This is because the net baryon diffusion current is mainly driven by the inward pointing spatial gradients of $\mu_B/T$, which act against local pressure gradients \cite{Denicol:2018wdp, Wu:QM2019}. The cross diffusion between the net baryon and net strangeness induces an oscillating distribution of the net strangeness current at late time \cite{Fotakis:2019nbq}. It will be interesting to see how this pattern is mapped to final state hadron correlations.
Because the net proton rapidity distribution is sensitive to both the initial state stopping and baryon diffusion \cite{Li:2018fow}, independent experimental observables are needed to disentangle these two effects. 
Recently, charge balance functions were proposed to independently constrain the charge diffusion constants \cite{Pratt:2019pnd}. A larger diffusion constant in the QGP medium leads to wider azimuthal distribution for the $K^+K^-$ and $p\bar{p}$ correlations.

\vspace{-5mm}
\section{The curious case of the shrinking QGP droplet: Challenges and opportunities in small systems}
\label{smallsystem}
\vspace{-2mm}

Collectivity has been studied systematically as a function of the collision system size at RHIC and LHC energies. As the size of a QGP droplet shrinks, the flow observables can reveal more information about the early-stage dynamics. At the same time, the increasing role of fluctuations stresses the ability of experiments to unambiguously identify flow signatures (see review by K. Gajdosova in this issue \cite{Gajdosova:QM2019}). On the theoretical side, small systems are also pushing the hydrodynamic framework to its limits. 

The IP-Glasma + MUSIC + UrQMD hybrid framework presented a remarkable success in describing the system size dependence of the flow measurements over more than two orders of magnitude in particle multiplicity \cite{Schenke:2020unx}. With a single set of model parameters, this theoretical framework can consistently describe particle production, radial and anisotropic flow observables, and multi-particle correlations from Pb+Pb to p+Pb collisions at the LHC and from Au+Au and to p+Au collisions at the top RHIC energy. That work demonstrated that the universal hydrodynamic response to collision geometry dominated the flow production in collisions with $dN^\mathrm{ch}/d\eta > 10$ at mid-rapidity. On the other hand, the correlation between $v_2$ and initial momentum anisotropy from the pre-hydrodynamic phase becomes stronger in the lower multiplicity collisions \cite{Schenke:2019pmk}. That work provided phenomenological evidence that the flow in low multiplicity collisions can elucidate the early dynamics of the collisions. A similar finding was presented by M. Luzum from a study of the correlation between the system's final elliptic flow and initial energy-momentum tensor \cite{Luzum:QM2019}.

The anisotropic flow in high energy p+p collisions still challenges our understanding of the underlying dynamics in these small systems. The computed $v_2\{2\}$ from the IP-Glasma hybrid framework increases as charged hadron multiplicity decreases \cite{Schenke:2020unx}, which is not seen in the flow measurements at the LHC. Comparing with results from \cite{Weller:2017tsr} suggests that the momentum anisotropy from the Glasma phase might be too strong in p+p collisions. 
Recent work pointed out that hydrodynamics introduces too strong non-linear interactions in p+p collisions, which results in a positive $C_2\{4\}$ opposite to experimental measurements \cite{Zhao:2020pty}. Elucidating the real dynamics in p+p collisions requires more theoretical progress on quantifying the contributions from non-hydrodynamic modes from both weakly-coupled effective transport descriptions \cite{Kurkela:2019kip} as well as strongly-coupled holographic approaches \cite{Buchel:2016cbj}. See the review by Y. Akamatsu for recent progress in understanding hydrodynamization and thermalization in heavy-ion collisions \cite{Akamatsu:2020lej}.

Finally, it is instructive to quantify to what extent small collision systems have been pushing the hydrodynamic framework to its limits. Conditions for non-linear causality bounds of second-order hydrodynamics were derived for radially expanding systems \cite{Floerchinger:2017cii} and for bulk viscosity for a general flow background \cite{Bemfica:2019cop}. These non-linear causality bounds set strong constraints on the maximal allowed viscous pressure in dynamical simulations. Because of the strong expansion rate and the consequent large negative bulk viscous pressure, the causality conditions in a typical p+A collision are $\sim$20\% closer to the bound than those in an A+A collision. The violent expansion in small systems also results in negative total (thermal + bulk viscous) pressure in a significant fraction of fluid cells. This may cause unstable cavitation inside the QGP \cite{Rajagopal:2009yw,Byres:2019xld}. The phenomenological impact of these negative pressure regions on final flow observables can be estimated by numerically regulating the size of bulk viscous pressure to be less than the thermal pressure. This modification leads to a sizable variation in the elliptic flow in 0-5\% p+Au collisions. For $p_T < 1.5$ GeV, this associated theoretical uncertainty is comparable to effects resulting from varying the second-order transport coefficients \cite{Schenke:2019pmk} . The situation in semi-peripheral A+A collisions is much better with a negligible effect on final flow observables. The current numerical regulation can only provide us with a rough estimate. Anisotropic hydrodynamics can provide us with more robust theory guidance on this issue \cite{Alqahtani:2017tnq,McNelis:2018jho}.

\vspace{-5mm}
\section{Conclusions}
\label{conclusion}
\vspace{-2mm}

Phenomenological studies combined with high precision flow measurements at RHIC and the LHC define the precision level in heavy-ion physics. Advanced statistical tools, such as Bayesian analysis, have become a standard approach to extract the QGP transport properties systematically. The next challenge lies in how to quantify the model uncertainty in the theoretical framework. Flow observables from the RHIC BES program and future FAIR/NICA experiments bring us to an era of full 3D dynamics. With the development of dynamical initialization schemes which interweave the 3D collision dynamics with fluid simulations, we are starting to quantify initial baryon stopping and study the collectivity of the QGP in a baryon-rich environment. This framework provides us with a reliable baseline to hunt for critical point signals in the upcoming RHIC BES II measurements.
Last but not least, flow in small systems offers a window to study the early-stage dynamics of QGP. Understanding the collective origin in small systems has been leading the state-of-the-art theory development in relativistic hydrodynamics.

\medskip

\noindent \textbf{Acknowledgments} The author thanks the organizers of Quark Matter 2019 for the invitation to give this plenary talk. The author thanks the JETSCAPE Collaboration for providing the preliminary results in Fig.~\ref{fig1} and for fruitful discussions with Y. Akamatsu, S. Bass, K. Eskola, K. Gajdosova, C. Gale, U. Heinz, A. Monnai, J. Noronha, J.~F. Paquet, and B. Schenke. This work was supported by the U.S. Department of Energy (DOE) under grant number DE-SC0013460 and within the framework of the Beam Energy Scan Theory (BEST) Topical Collaboration.




\vspace{-4mm}
\bibliographystyle{elsarticle-num-1}
\bibliography{references}

\begin{thebibliography}{10}
\expandafter\ifx\csname url\endcsname\relax
  \def\url#1{\texttt{#1}}\fi
\expandafter\ifx\csname urlprefix\endcsname\relax\def\urlprefix{URL }\fi
\expandafter\ifx\csname href\endcsname\relax
  \def\href#1#2{#2} \def\path#1{#1}\fi

\bibitem{Ollitrault:1992bk}
J.-Y. Ollitrault, {Anisotropy as a signature of transverse collective flow},
  Phys. Rev. D46 (1992) 229--245.

\bibitem{Khachatryan:2015waa}
V.~Khachatryan, et~al., {Evidence for collective multiparticle correlations in
  p-Pb collisions}, Phys. Rev. Lett. 115~(1) (2015) 012301.
\newblock \href {http://arxiv.org/abs/1502.05382} {\path{ arXiv:1502.05382}}.

\bibitem{ALICE:2016kpq}
J.~Adam, et~al., {Correlated event-by-event fluctuations of flow harmonics in
  Pb-Pb collisions at $\sqrt{s_{_{\rm NN}}}=2.76$ TeV}, Phys. Rev. Lett. 117
  (2016) 182301.
\newblock \href {http://arxiv.org/abs/1604.07663} {\path{ arXiv:1604.07663}}.

\bibitem{Aad:2014fla}
G.~Aad, et~al., {Measurement of event-plane correlations in
  $\sqrt{s_{NN}}=2.76$ TeV lead-lead collisions with the ATLAS detector}, Phys.
  Rev. C90~(2) (2014) 024905.
\newblock \href {http://arxiv.org/abs/1403.0489} {\path{ arXiv:1403.0489}}.

\bibitem{Aad:2013xma}
G.~Aad, et~al., {Measurement of the distributions of event-by-event flow
  harmonics in lead-lead collisions at = 2.76 TeV with the ATLAS detector at
  the LHC}, JHEP 11 (2013) 183.
\newblock \href {http://arxiv.org/abs/1305.2942} {\path{ arXiv:1305.2942}}.

\bibitem{Heinz:2013th}
U.~Heinz, R.~Snellings, {Collective flow and viscosity in relativistic
  heavy-ion collisions}, Ann. Rev. Nucl. Part. Sci. 63 (2013) 123--151.
\newblock \href {http://arxiv.org/abs/1301.2826} {\path{ arXiv:1301.2826}}.

\bibitem{Gale:2013da}
C.~Gale, S.~Jeon, B.~Schenke, {Hydrodynamic modeling of heavy-ion collisions},
  Int. J. Mod. Phys. A28 (2013) 1340011.

\bibitem{Shen:2016odt}
C.~Shen, {Electromagnetic radiation from QCD matter: theory overview}, Nucl.
  Phys. A956 (2016) 184--191.
\newblock \href {http://arxiv.org/abs/1601.02563} {\path{ arXiv:1601.02563}}.

\bibitem{Putschke:2019yrg}
J.~H. Putschke, et~al., {The JETSCAPE framework}\href
  {http://arxiv.org/abs/1903.07706} {\path{ arXiv:1903.07706}}.

\bibitem{Israel:1979wp}
W.~Israel, J.~M. Stewart, {Transient relativistic thermodynamics and kinetic
  theory}, Annals Phys. 118 (1979) 341--372.

\bibitem{Baier:2007ix}
R.~Baier, P.~Romatschke, D.~T. Son, A.~O. Starinets, M.~A. Stephanov,
  {Relativistic viscous hydrodynamics, conformal invariance, and holography},
  JHEP 04 (2008) 100.
\newblock \href {http://arxiv.org/abs/0712.2451} {\path{ arXiv:0712.2451}}.

\bibitem{Denicol:2012cn}
G.~S. Denicol, H.~Niemi, E.~Molnar, D.~H. Rischke, {Derivation of transient
  relativistic fluid dynamics from the Boltzmann equation}, Phys. Rev. D85
  (2012) 114047, [Erratum: Phys. Rev. D91, no.3, 039902 (2015)].
\newblock \href {http://arxiv.org/abs/1202.4551} {\path{ arXiv:1202.4551}}.

\bibitem{Strickland:2014pga}
M.~Strickland, {Anisotropic hydrodynamics: three lectures}, Acta Phys. Polon.
  B45~(12) (2014) 2355--2394.
\newblock \href {http://arxiv.org/abs/1410.5786} {\path{ arXiv:1410.5786}}.

\bibitem{Jeon:2015dfa}
S.~Jeon, U.~Heinz, {Introduction to Hydrodynamics}, Int. J. Mod. Phys. E24~(10)
  (2015) 1530010.
\newblock \href {http://arxiv.org/abs/1503.03931} {\path{ arXiv:1503.03931}}.

\bibitem{Bass:1998ca}
S.~A. Bass, et~al., {Microscopic models for ultrarelativistic heavy ion
  collisions}, Prog. Part. Nucl. Phys. 41 (1998) 255--369, [Prog. Part. Nucl.
  Phys.41,225(1998)].
\newblock \href {http://arxiv.org/abs/nucl-th/9803035} {\path{
  arXiv:nucl-th/9803035}}.

\bibitem{Bleicher:1999xi}
M.~Bleicher, et~al., {Relativistic hadron hadron collisions in the
  ultrarelativistic quantum molecular dynamics model}, J. Phys. G25 (1999)
  1859--1896.
\newblock \href {http://arxiv.org/abs/hep-ph/9909407} {\path{
  arXiv:hep-ph/9909407}}.

\bibitem{Nara:1999dz}
Y.~Nara, N.~Otuka, A.~Ohnishi, K.~Niita, S.~Chiba, {Study of relativistic
  nuclear collisions at AGS energies from p + Be to Au + Au with hadronic
  cascade model}, Phys. Rev. C61 (2000) 024901.
\newblock \href {http://arxiv.org/abs/nucl-th/9904059} {\path{
  arXiv:nucl-th/9904059}}.

\bibitem{Weil:2016zrk}
J.~Weil, et~al., {Particle production and equilibrium properties within a new
  hadron transport approach for heavy-ion collisions}, Phys. Rev. C94~(5)
  (2016) 054905.
\newblock \href {http://arxiv.org/abs/1606.06642} {\path{ arXiv:1606.06642}}.

\bibitem{Gale:QM2019}
See contribution from C. Gale in this issue.

\bibitem{Schenke:2012wb}
B.~Schenke, P.~Tribedy, R.~Venugopalan, {Fluctuating Glasma initial conditions
  and flow in heavy ion collisions}, Phys. Rev. Lett. 108 (2012) 252301.
\newblock \href {http://arxiv.org/abs/1202.6646} {\path{ arXiv:1202.6646}}.

\bibitem{Kurkela:2018wud}
A.~Kurkela, A.~Mazeliauskas, J.-F. Paquet, S.~Schlichting, D.~Teaney, {Matching
  the nonequilibrium initial stage of heavy ion collisions to hydrodynamics
  with QCD kinetic theory}, Phys. Rev. Lett. 122~(12) (2019) 122302.
\newblock \href {http://arxiv.org/abs/1805.01604} {\path{ arXiv:1805.01604}}.

\bibitem{Paquet:2015lta}
J.-F. Paquet, C.~Shen, G.~S. Denicol, M.~Luzum, B.~Schenke, S.~Jeon, C.~Gale,
  {Production of photons in relativistic heavy-ion collisions}, Phys. Rev.
  C93~(4) (2016) 044906.
\newblock \href {http://arxiv.org/abs/1509.06738} {\path{ arXiv:1509.06738}}.

\bibitem{Schenke:2020unx}
B.~Schenke, C.~Shen, P.~Tribedy, {Bulk properties and multi-particle
  correlations in large and small systems}.
\newblock \href {http://arxiv.org/abs/2001.09949} {\path{ arXiv:2001.09949}}.

\bibitem{Liu:2015nwa}
J.~Liu, C.~Shen, U.~Heinz, {Pre-equilibrium evolution effects on heavy-ion
  collision observables}, Phys. Rev. C91~(6) (2015) 064906, [Erratum: Phys.
  Rev. C92, 049904 (2015)].
\newblock \href {http://arxiv.org/abs/1504.02160} {\path{ arXiv:1504.02160}}.

\bibitem{Pratt:2015zsa}
S.~Pratt, E.~Sangaline, P.~Sorensen, H.~Wang, {Constraining the Eq. of State of
  super-hadronic matter from heavy-ion collisions}, Phys. Rev. Lett. 114 (2015)
  202301.
\newblock \href {http://arxiv.org/abs/1501.04042} {\path{ arXiv:1501.04042}}.

\bibitem{Bernhard:2019bmu}
J.~E. Bernhard, J.~S. Moreland, S.~A. Bass, {Bayesian estimation of the
  specific shear and bulk viscosity of quark–gluon plasma}, Nature Phys.
  15~(11) (2019) 1113--1117.

\bibitem{Paquet:QM2019}
See contribution from J.F. Paquet in this issue.

\bibitem{Gale:2012rq}
C.~Gale, S.~Jeon, B.~Schenke, P.~Tribedy, R.~Venugopalan, {Event-by-event
  anisotropic flow in heavy-ion collisions from combined Yang-Mills and viscous
  fluid dynamics}, Phys. Rev. Lett. 110~(1) (2013) 012302.
\newblock \href {http://arxiv.org/abs/1209.6330} {\path{ arXiv:1209.6330}}.

\bibitem{Ryu:2015vwa}
S.~Ryu, J.~F. Paquet, C.~Shen, G.~S. Denicol, B.~Schenke, S.~Jeon, C.~Gale,
  {Importance of the bulk viscosity of QCD in ultrarelativistic heavy-ion
  collisions}, Phys. Rev. Lett. 115~(13) (2015) 132301.
\newblock \href {http://arxiv.org/abs/1502.01675} {\path{ arXiv:1502.01675}}.

\bibitem{Schenke:2018fci}
B.~Schenke, C.~Shen, P.~Tribedy, {Features of the IP-Glasma}, Nucl. Phys. A982
  (2019) 435--438.
\newblock \href {http://arxiv.org/abs/1807.05205} {\path{ arXiv:1807.05205}}.

\bibitem{Niemi:2015qia}
H.~Niemi, K.~J. Eskola, R.~Paatelainen, {Event-by-event fluctuations in a
  perturbative QCD + saturation + hydrodynamics model: Determining QCD matter
  shear viscosity in ultrarelativistic heavy-ion collisions}, Phys. Rev.
  C93~(2) (2016) 024907.

\bibitem{Bernhard:2016tnd}
J.~E. Bernhard, J.~S. Moreland, S.~A. Bass, J.~Liu, U.~Heinz, {Applying
  Bayesian parameter estimation to relativistic heavy-ion collisions:
  simultaneous characterization of the initial state and quark-gluon plasma
  medium}, Phys. Rev. C94~(2) (2016) 024907.

\bibitem{Borsanyi:2013bia}
S.~Borsanyi, Z.~Fodor, C.~Hoelbling, S.~D. Katz, S.~Krieg, K.~K. Szabo, {Full
  result for the QCD equation of state with 2+1 flavors}, Phys. Lett. B730
  (2014) 99--104.
\newblock \href {http://arxiv.org/abs/1309.5258} {\path{ arXiv:1309.5258}}.

\bibitem{Bazavov:2014pvz}
A.~Bazavov, et~al., {Equation of state in ( 2+1 )-flavor QCD}, Phys. Rev. D90
  (2014) 094503.
\newblock \href {http://arxiv.org/abs/1407.6387} {\path{ arXiv:1407.6387}}.

\bibitem{Mazeliauskas:2015efa}
A.~Mazeliauskas, D.~Teaney, {Fluctuations of harmonic and radial flow in heavy
  ion collisions with principal components}, Phys. Rev. C93~(2) (2016) 024913.
\newblock \href {http://arxiv.org/abs/1509.07492} {\path{ arXiv:1509.07492}}.

\bibitem{Hippert:2019swu}
M.~Hippert, D.~Dobrigkeit~Chinellato, M.~Luzum, J.~Noronha, T.~Nunes~da Silva,
  J.~Takahashi, {Measuring momentum-dependent flow fluctuations in heavy-ion
  collisions}\href {http://arxiv.org/abs/1906.08915} {\path{
  arXiv:1906.08915}}.

\bibitem{Gardim:2019iah}
F.~G. Gardim, F.~Grassi, P.~Ishida, M.~Luzum, J.-Y. Ollitrault,
  {$p_T$-dependent particle number fluctuations from principal component
  analyses in hydrodynamic simulations of heavy-ion collisions}, Phys. Rev.
  C100~(5) (2019) 054905.
\newblock \href {http://arxiv.org/abs/1906.03045} {\path{ arXiv:1906.03045}}.

\bibitem{Pang:QM2019}
See contribution from L. Pang in this issue.

\bibitem{Shen:2017bsr}
C.~Shen, B.~Schenke, {Dynamical initial state model for relativistic heavy-ion
  collisions}, Phys. Rev. C97~(2) (2018) 024907.

\bibitem{Monnai:2019hkn}
A.~Monnai, B.~Schenke, C.~Shen, {Equation of state at finite densities for QCD
  matter in nuclear collisions}, Phys. Rev. C100~(2) (2019) 024907.
\newblock \href {http://arxiv.org/abs/1902.05095} {\path{ arXiv:1902.05095}}.

\bibitem{Adamczyk:2016exq}
L.~Adamczyk, et~al., {Beam energy dependence of the third harmonic of azimuthal
  correlations in Au+Au collisions at RHIC}, Phys. Rev. Lett. 116~(11) (2016)
  112302.
\newblock \href {http://arxiv.org/abs/1601.01999} {\path{ arXiv:1601.01999}}.

\bibitem{Bazavov:2018mes}
A.~Bazavov, et~al., {Chiral crossover in QCD at zero and non-zero chemical
  potentials}, Phys. Lett. B795 (2019) 15--21.

\bibitem{Andronic:2017pug}
A.~Andronic, P.~Braun-Munzinger, K.~Redlich, J.~Stachel, {Decoding the phase
  structure of QCD via particle production at high energy}, Nature 561~(7723)
  (2018) 321--330.
\newblock \href {http://arxiv.org/abs/1710.09425} {\path{ arXiv:1710.09425}}.

\bibitem{Adamczyk:2017iwn}
L.~Adamczyk, et~al., {Bulk properties of the medium produced in relativistic
  heavy-ion Collisions from the beam energy scan program}, Phys. Rev. C96~(4)
  (2017) 044904.
\newblock \href {http://arxiv.org/abs/1701.07065} {\path{ arXiv:1701.07065}}.

\bibitem{Bialas:2016epd}
A.~Bialas, A.~Bzdak, V.~Koch, {Stopped nucleons in configuration space}, Acta
  Phys. Polon. B49 (2018) 103.
\newblock \href {http://arxiv.org/abs/1608.07041} {\path{ arXiv:1608.07041}}.

\bibitem{Karpenko:2015xea}
I.~A. Karpenko, P.~Huovinen, H.~Petersen, M.~Bleicher, {Estimation of the shear
  viscosity at finite net-baryon density from $A+A$ collision data at
  $\sqrt{s_\mathrm{NN}} = 7.7-200$ GeV}, Phys. Rev. C91~(6) (2015) 064901.
\newblock \href {http://arxiv.org/abs/1502.01978} {\path{ arXiv:1502.01978}}.

\bibitem{Du:2018mpf}
L.~Du, U.~Heinz, G.~Vujanovic, {Hybrid model with dynamical sources for
  heavy-ion collisions at BES energies}, Nucl. Phys. A982 (2019) 407--410.
\newblock \href {http://arxiv.org/abs/1807.04721} {\path{ arXiv:1807.04721}}.

\bibitem{Li:2018ini}
M.~Li, J.~I. Kapusta, {Large baryon densities achievable in high energy heavy
  ion collisions outside the central rapidity region}, Phys. Rev. C99~(1)
  (2019) 014906.
\newblock \href {http://arxiv.org/abs/1808.05751} {\path{ arXiv:1808.05751}}.

\bibitem{McLerran:2018avb}
L.~D. McLerran, S.~Schlichting, S.~Sen, {Spacetime picture of baryon stopping
  in the color-glass condensate}, Phys. Rev. D99~(7) (2019) 074009.
\newblock \href {http://arxiv.org/abs/1811.04089} {\path{ arXiv:1811.04089}}.

\bibitem{Attems:2018gou}
M.~Attems, Y.~Bea, J.~Casalderrey-Solana, D.~Mateos, M.~Triana, M.~Zilhão,
  {Holographic collisions across a phase transition}, Phys. Rev. Lett. 121~(26)
  (2018) 261601.
\newblock \href {http://arxiv.org/abs/1807.05175} {\path{ arXiv:1807.05175}}.

\bibitem{Okai:2017ofp}
M.~Okai, K.~Kawaguchi, Y.~Tachibana, T.~Hirano, {New approach to initializing
  hydrodynamic fields and mini-jet propagation in quark-gluon fluids}, Phys.
  Rev. C95~(5) (2017) 054914.
\newblock \href {http://arxiv.org/abs/1702.07541} {\path{ arXiv:1702.07541}}.

\bibitem{Shen:2017ruz}
C.~Shen, G.~Denicol, C.~Gale, S.~Jeon, A.~Monnai, B.~Schenke, {A hybrid
  approach to relativistic heavy-ion collisions at the RHIC BES energies},
  Nucl. Phys. A967 (2017) 796--799.
\newblock \href {http://arxiv.org/abs/1704.04109} {\path{ arXiv:1704.04109}}.

\bibitem{Kanakubo:2019ogh}
Y.~Kanakubo, Y.~Tachibana, T.~Hirano, {Unified description of hadron chemistry
  from dynamical core--corona initialization}\href
  {http://arxiv.org/abs/1910.10556} {\path{ arXiv:1910.10556}}.

\bibitem{Ratti:2018ksb}
C.~Ratti, {Lattice QCD and heavy ion collisions: a review of recent progress},
  Rept. Prog. Phys. 81~(8) (2018) 084301.

\bibitem{Noronha-Hostler:2019ayj}
J.~Noronha-Hostler, P.~Parotto, C.~Ratti, J.~M. Stafford, {Lattice-based
  equation of state at finite baryon number, electric charge and strangeness
  chemical potentials}, Phys. Rev. C100~(6) (2019) 064910.
\newblock \href {http://arxiv.org/abs/1902.06723} {\path{ arXiv:1902.06723}}.

\bibitem{Parotto:2018pwx}
P.~Parotto, M.~Bluhm, D.~Mroczek, M.~Nahrgang, J.~Noronha-Hostler,
  K.~Rajagopal, C.~Ratti, T.~Schäfer, M.~Stephanov, {Lattice-QCD-based
  equation of state with a critical point}\href
  {http://arxiv.org/abs/1805.05249} {\path{ arXiv:1805.05249}}.

\bibitem{Shen:2018pty}
C.~Shen, B.~Schenke, {Dynamical initialization and hydrodynamic modeling of
  relativistic heavy-ion collisions}, Nucl. Phys. A982 (2019) 411--414.
\newblock \href {http://arxiv.org/abs/1807.05141} {\path{ arXiv:1807.05141}}.

\bibitem{Adamczyk:2017hdl}
L.~Adamczyk, et~al., {Harmonic decomposition of three-particle azimuthal
  correlations at energies available at the BNL Relativistic Heavy Ion
  Collider}, Phys. Rev. C98~(3) (2018) 034918.
\newblock \href {http://arxiv.org/abs/1701.06496} {\path{ arXiv:1701.06496}}.

\bibitem{Bozek:2017qir}
P.~Bozek, W.~Broniowski, {Longitudinal decorrelation measures of flow magnitude
  and event-plane angles in ultrarelativistic nuclear collisions}, Phys. Rev.
  C97~(3) (2018) 034913.
\newblock \href {http://arxiv.org/abs/1711.03325} {\path{ arXiv:1711.03325}}.

\bibitem{Li:2019eni}
H.~Li, L.~Yan, {Pseudorapidity dependent hydrodynamic response in heavy-ion
  collisions}\href {http://arxiv.org/abs/1907.10854} {\path{
  arXiv:1907.10854}}.

\bibitem{McDonald:2020oyf}
S.~McDonald, S.~Jeon, C.~Gale, {Exploring longitudinal oservables with 3+1D
  IP-Glasma}, 2020.
\newblock \href {http://arxiv.org/abs/2001.08636} {\path{ arXiv:2001.08636}}.

\bibitem{Sakai:QM2019}
See contribution from A. Sakai in this issue.

\bibitem{Nie:QM2019}
See contribution from M. Nie in this issue.

\bibitem{Martinez:2019jbu}
M.~Martinez, M.~D. Sievert, D.~E. Wertepny, J.~Noronha-Hostler, {Initial state
  fluctuations of QCD conserved charges in heavy-ion collisions}\href
  {http://arxiv.org/abs/1911.10272} {\path{ arXiv:1911.10272}}.

\bibitem{Martinez:2019rlp}
M.~Martinez, M.~D. Sievert, D.~E. Wertepny, J.~Noronha-Hostler, {Toward initial
  conditions of conserved charges part II: The ICCING monte carlo
  algorithm}\href {http://arxiv.org/abs/1911.12454} {\path{ arXiv:1911.12454}}.

\bibitem{Pratt:2012dz}
S.~Pratt, {Identifying the Charge Carriers of the Quark-Gluon Plasma}, Phys.
  Rev. Lett. 108 (2012) 212301.
\newblock \href {http://arxiv.org/abs/1203.4578} {\path{ arXiv:1203.4578}}.

\bibitem{Greif:2017byw}
M.~Greif, J.~A. Fotakis, G.~S. Denicol, C.~Greiner, {Diffusion of conserved
  charges in relativistic heavy ion collisions}, Phys. Rev. Lett. 120~(24)
  (2018) 242301.
\newblock \href {http://arxiv.org/abs/1711.08680} {\path{ arXiv:1711.08680}}.

\bibitem{Fotakis:2019nbq}
J.~A. Fotakis, M.~Greif, G.~Denicol, H.~Niemi, C.~Greiner, {Diffusion processes
  involving multiple conserved charges: a first study from kinetic theory and
  implications to the fluid-dynamical modeling of heavy ion collisions}\href
  {http://arxiv.org/abs/1912.09103} {\path{ arXiv:1912.09103}}.

\bibitem{Denicol:2018wdp}
G.~S. Denicol, C.~Gale, S.~Jeon, A.~Monnai, B.~Schenke, C.~Shen, {Net baryon
  diffusion in fluid dynamic simulations of relativistic heavy-ion collisions},
  Phys. Rev. C98~(3) (2018) 034916.
\newblock \href {http://arxiv.org/abs/1804.10557} {\path{ arXiv:1804.10557}}.

\bibitem{Wu:QM2019}
See contribution from X. Wu in this issue.

\bibitem{Li:2018fow}
M.~Li, C.~Shen, {Longitudinal dynamics of high baryon density matter in high
  energy heavy-ion collisions}, Phys. Rev. C98~(6) (2018) 064908.
\newblock \href {http://arxiv.org/abs/1809.04034} {\path{ arXiv:1809.04034}}.

\bibitem{Pratt:2019pnd}
S.~Pratt, C.~Plumberg, {Determining the Diffusivity for Light Quarks from
  Experiment}\href {http://arxiv.org/abs/1904.11459} {\path{
  arXiv:1904.11459}}.

\bibitem{Gajdosova:QM2019}
See contribution from K. Gajdosova in this issue.

\bibitem{Schenke:2019pmk}
B.~Schenke, C.~Shen, P.~Tribedy, {Hybrid Color Glass Condensate and
  hydrodynamic description of the Relativistic Heavy Ion Collider small system
  scan}\href {http://arxiv.org/abs/1908.06212} {\path{ arXiv:1908.06212}}.

\bibitem{Luzum:QM2019}
See contribution from M. Luzum in this issue.

\bibitem{Weller:2017tsr}
R.~D. Weller, P.~Romatschke, {One fluid to rule them all: viscous hydrodynamic
  description of event-by-event central p+p, p+Pb and Pb+Pb collisions at
  $\sqrt{s}=5.02$ TeV}, Phys. Lett. B774 (2017) 351--356.
\newblock \href {http://arxiv.org/abs/1701.07145} {\path{ arXiv:1701.07145}}.

\bibitem{Zhao:2020pty}
W.~Zhao, Y.~Zhou, K.~Murase, H.~Song, {Searching for small droplets of
  hydrodynamic fluid in proton--proton collisions at the LHC}\href
  {http://arxiv.org/abs/2001.06742} {\path{ arXiv:2001.06742}}.

\bibitem{Kurkela:2019kip}
A.~Kurkela, U.~A. Wiedemann, B.~Wu, {Flow in AA and pA as an interplay of
  fluid-like and non-fluid like excitations}, Eur. Phys. J. C79~(11) (2019)
  965.
\newblock \href {http://arxiv.org/abs/1905.05139} {\path{ arXiv:1905.05139}}.

\bibitem{Buchel:2016cbj}
A.~Buchel, M.~P. Heller, J.~Noronha, {Entropy production, hydrodynamics, and
  resurgence in the primordial Quark-Gluon Plasma from holography}, Phys. Rev.
  D94~(10) (2016) 106011.
\newblock \href {http://arxiv.org/abs/1603.05344} {\path{ arXiv:1603.05344}}.

\bibitem{Akamatsu:2020lej}
Y.~Akamatsu, {Approach to thermalization and hydrodynamics}, in: {28th
  International Conference on Ultrarelativistic Nucleus-Nucleus Collisions
  (Quark Matter 2019) Wuhan, China, November 4-9, 2019}, 2020.
\newblock \href {http://arxiv.org/abs/2001.01429} {\path{ arXiv:2001.01429}}.

\bibitem{Floerchinger:2017cii}
S.~Floerchinger, E.~Grossi, {Causality of fluid dynamics for high-energy
  nuclear collisions}, JHEP 08 (2018) 186.

\bibitem{Bemfica:2019cop}
F.~S. Bemfica, M.~M. Disconzi, J.~Noronha, {Causality of the
  Einstein-Israel-Stewart theory with bulk viscosity}, Phys. Rev. Lett.
  122~(22) (2019) 221602.
\newblock \href {http://arxiv.org/abs/1901.06701} {\path{ arXiv:1901.06701}}.

\bibitem{Rajagopal:2009yw}
K.~Rajagopal, N.~Tripuraneni, {Bulk viscosity and cavitation in boost-invariant
  hydrodynamic expansion}, JHEP 03 (2010) 018.

\bibitem{Byres:2019xld}
M.~Byres, S.~H. Lim, C.~McGinn, J.~Ouellette, J.~L. Nagle, {The Skinny on Bulk
  Viscosity and Cavitation in Heavy Ion Collisions}\href
  {http://arxiv.org/abs/1910.12930} {\path{ arXiv:1910.12930}}.

\bibitem{Alqahtani:2017tnq}
M.~Alqahtani, M.~Nopoush, R.~Ryblewski, M.~Strickland, {Anisotropic
  hydrodynamic modeling of 2.76 TeV Pb-Pb collisions}, Phys. Rev. C96~(4)
  (2017) 044910.
\newblock \href {http://arxiv.org/abs/1705.10191} {\path{ arXiv:1705.10191}}.

\bibitem{McNelis:2018jho}
M.~McNelis, D.~Bazow, U.~Heinz, {(3+1)-dimensional anisotropic fluid dynamics
  with a lattice QCD equation of state}, Phys. Rev. C97~(5) (2018) 054912.
\newblock \href {http://arxiv.org/abs/1803.01810} {\path{ arXiv:1803.01810}}.

\end{thebibliography}







\end{document}